%% file: conference_101719.tex
\documentclass[conference]{IEEEtran}
\IEEEoverridecommandlockouts
\usepackage[english]{babel}
\usepackage{blindtext}
\usepackage{tikz,pgfplots}
\usetikzlibrary{calc}
\usepackage{tkz-euclide}
\usepackage{bm}
\pgfplotsset{compat=1.5}
\usepgfplotslibrary{polar}
\usepackage{epstopdf} 
\usepackage{amsmath}
\usepackage{wrapfig}
\usepackage{verbatim}
\usepackage{mathrsfs}
\usepackage{amssym  b}
\usepackage{lipsum}  
\usepackage{acronym}
\usepackage{graphicx}
\usepackage{textcomp}
\usepackage{xcolor}
\usepackage{mathtools}
\usepackage{color,soul}
\usepackage{graphicx}
\usepackage{footnote}
\usepackage{booktabs}
\usepackage{multirow}
\usepackage{tikz}
\usepackage[font={small}]{caption}
\usepackage[font={small}]{subcaption}
\usepackage{algorithm}
\usepackage{algorithmic}
\usepackage{fancyhdr}

\usepackage{caption}
\usepackage{subcaption}
\usepackage{cleveref}
\usepackage{url}

\newcommand{\eq}[1]{Eq.~\eqref{#1}}
\newcommand{\fig}[1]{Fig.~\ref{#1}}
\newcommand{\tab}[1]{Tab.~\ref{#1}}
\newcommand{\secref}[1]{Section~\ref{#1}}

\input{ACRONYMS.tex}

\begin{document}

\title{Learned Spike Encoding of the Channel Response for Low-Power Environment Sensing}

\author{Eleonora Cicciarella$^{*\dag}$, Riccardo Mazzieri$^{\dag}$, Jacopo Pegoraro$^{\dag}$, Michele Rossi$^{\mathsection \dag}$
\thanks{$^{*}$Corresponding author \texttt{eleonora.cicciarella@phd.unipd.it}.
$^{\dag }$These authors are with the University of Padova, Department of Information Engineering.
$^{\mathsection}$This author is with the University of Padova, Department of Mathematics ``Tullio Levi-Civita''. 

This work was partially supported by the European Union under the Italian National Recovery and Resilience Plan (NRRP) of NextGenerationEU, partnership on “Telecommunications of the Future” (PE0000001 - program “RESTART”).}}

\IEEEoverridecommandlockouts
\IEEEaftertitletext{\vspace{-1\baselineskip}}
\maketitle

\begin{abstract}
Radio Frequency (RF) sensing holds the potential for enabling pervasive monitoring applications. However, modern sensing algorithms imply complex operations, which clash with the energy-constrained nature of edge sensing devices. This calls for the development of new processing and \textit{learning} techniques that can strike a suitable balance between performance and energy efficiency. 
Spiking Neural Networks (SNNs) have recently emerged as an energy-efficient alternative to conventional neural networks for edge computing applications. They process information in the form of {\it sparse} binary spike trains, thus potentially reducing energy consumption by several orders of magnitude. Their fruitful use for RF signal processing critically depends on the representation of RF signals in the form of spike signals. We underline that existing \textit{spike encoding} algorithms to do so generally produce inaccurate signal representations and dense (i.e., inefficient) spike trains.

In this work, we propose a lightweight neural architecture that learns a \textit{tailored} spike encoding representations of RF channel responses by jointly reconstructing the input and its spectral content. By leveraging a tunable regularization term, our approach enables fine-grained control over the performance-energy trade-off of the system. Our numerical results show that the proposed method outperforms existing encoding algorithms in terms of reconstruction error and sparsity of the obtained spike encodings.
\end{abstract}


\section{Introduction}

\ac{rf} sensing has gathered a huge interest as a privacy-preserving and ubiquitous alternative to cameras for environment monitoring, pervasive healthcare, and touchless human-machine interaction, among others~\cite{liu2019wireless, shastri2022review}.
Depending on the adopted technology, existing \ac{rf} sensing systems can use dedicated commercial radar sensors adopting \ac{fmcw} or \ac{uwb} signals~\cite{shastri2022review}, reuse the signals transmitted by existing Wi-Fi routers or by cellular base stations~\cite{ma2019wifi} (termed \ac{isac}). 
\ac{rf} sensing entails the estimation of the wireless channel, as the latter embeds information about the location and movement of objects and people in the physical surroundings. Due to the complexity of such channel, \ac{dl} solutions are commonly employed to enable advanced sensing applications such as target recognition, movement analysis, and human activity recognition~\cite{pegoraro2023rapid, ma2019wifi}. 
However, \ac{dl} is well-known to be \textit{energy-hungry} and, in turn, unsuited for implementation on resource-constrained edge devices and routers, where the \ac{rf} data are collected~\cite{lahmer2022energy}. This is further exacerbated by the extremely high sampling rates of modern analog-to-digital converters, which cause \ac{rf} sensors to generate a huge amount of data compared to other sensing modalities. 
Hence, a widespread adoption of \ac{rf} sensing, and \ac{isac} in particular, raises concerns regarding the large carbon footprint of the resulting sensor networks.

A possible solution to tackle the energy efficiency problem of \ac{rf} sensing at the network edge has been identified in replacing standard \ac{dl} architectures with \acp{snn}~\cite{eshraghian2023training}. \acp{snn} closely mimic the natural functioning of the brain, which is a strikingly energy-efficient processor operating within \mbox{$\sim 12-20$~W} of power. They encode information into event-based \textit{spike trains}, which can be represented as sparse time-domain binary signals~\cite{eshraghian2023training}.
Their energy consumption is directly related to the number of processed spikes and can be orders of magnitude smaller than than of standard GPU accelerators~\cite{frenkel20180}. As such, the way in which the input (channel) information is encoded into spikes is key in determining the trade-off between \textit{performance} on the learning task and \textit{sparsity} of the resulting spike train, which is tied to the energy consumption.
However, most available works on \acp{snn} for \ac{rf} sensing overlook the spike encoding step and rely on simple, general purpose techniques~\cite{safa2021improving, safa2021use}. Specifically, these approaches perform standard radar signal processing operations to estimate wireless channel features and then feed them to threshold- or moving-window-based encoding methods~\cite{petro2019selection}. These are not tailored for \ac{rf} sensing data, and offer limited control on the sparsity of the spike trains. Other preliminary works use the raw \ac{rf} samples as the \ac{snn} input~\cite{ arsalan2022spiking, chen2022neuromorphic}, thus avoiding the use of encoding algorithms.  
Notably, \cite{chen2022neuromorphic} proposes an \ac{snn}-based \ac{isac} system using short \ac{uwb} pulses, which are naturally represented as spikes after receiver sampling. These solutions achieve a better integration of the \ac{snn} into the \ac{rf} signal processing chain, but do not address the minimization of the energy consumption.

In this work, we propose a new approach to gain control over the performance-energy trade-off in \ac{rf} sensing with \acp{snn}. Our key insight is that channel representations in \ac{rf} sensing can be cast under a \textit{sum of sinusoids} model, in which the main features of interest are the frequency components and their amplitudes. 
We propose a lightweight neural architecture that learns a \textit{tailored} spike encoding for \ac{rf} channel responses (treated as input signals), preserving their spectral content. This architecture is trained to jointly optimize the fidelity to the original spectrum (hence the sensing performance) and the energy consumption, by promoting the sparsity of the encoded spike train. 

The main contributions of this work are summarized next.

\textbf{1)} We identify a sum of sinusoids signal model to represent the channel response in most \ac{rf} sensing tasks of interest (like, e.g., range, Doppler, and \ac{aoa} estimation). Hence, we leverage this model to formulate a spike encoding learning mechanism that achieves a desirable trade-off between fidelity to the original channel properties and energy consumption. This is achieved by minimizing the number of spikes in the encoding.

\textbf{2)} We design and validate a \ac{cae} that leverages a \ac{ste} to obtain a spike train feature encoding. Two output branches are used. One reconstructs the original channel representation, pushing the spike train to accurately represent the input. The other is a \ac{snn} performing a regression on the frequency components of the channel representation. To the best of our knowledge, this is the first encoding method to \textit{learn} spike patterns from \ac{rf} sensing data.

\textbf{3)} We train the proposed network on a synthetic dataset, showing its excellent capabilities to effectively encode \ac{rf} channel responses. Specifically, our method encodes channel features (and its spectral content in particular) by achieving spike trains that are more than $70\%$ sparser than those obtained from existing encoding algorithms.

\section{Background and signal model}\label{sec:background}

In both radar and \ac{isac}, sensing is performed by estimating the wireless propagation channel, which embeds information about the location and movement of objects and people in the surroundings. 
The key motivation behind our learned encoding design and signal model (see \secref{sec:signal-model}) lies in the common representation of the channel response for both radar and \ac{isac} processing, which takes a \textit{sum of sinusoids} form~\cite{patole2017automotive, pegoraro2023rapid, ma2019wifi}. For example, \ac{fmcw} radars estimate distance, Doppler speed, and \ac{aoa} of the sensing targets by processing the echo of a probe waveform, obtaining the so-called intermediate frequency signal~\cite{patole2017automotive}. This has a complex sinusoidal shape in the delay, Doppler, and spatial domains, which allows sensing parameters estimation via 3D-\ac{dft}. Although different waveforms and processing are employed, also \ac{isac} systems based on \ac{ofdm} or single-carrier transmission obtain similar channel representations, as detailed, e.g., in \cite{pegoraro2023rapid, ma2019wifi}. Hence, to design low-power \ac{rf} sensing technology based on \acp{snn} it is fundamentally important to efficiently encode sum of sinusoids channels into spike trains that: (i)~preserve the \textit{frequency content} of the channel response, as this contains the sensing parameters of interest and (ii)~contain as \textit{few spikes} as possible, to minimize the energy consumption.

In the following, we introduce the signal model representing the channel response as a sum of sinusoids and provide useful background on \acp{snn}.

\subsection{Signal model}\label{sec:signal-model}

As anticipated above, a sum of sinusoids model is considered to represent the channel response in the delay, Doppler, or spatial domains. The number of sinusoidal components is denoted by $M$, which correspond to the channel reflectors. Moreover, $a_m, \phi_m, f_m$ respectively represent the amplitude (accounting for the propagation and reflection losses), the phase, and the frequency of the $m$-th component. The sampling time is denoted by $T$ and a processing window of $K$ subsequent samples is considered, indexed by $k=0,\dots, K-1$. The signal model is written as 
\begin{equation}
\label{eq:signal_model}
    x[k] \triangleq x(kT) = \sum_{m=1}^M a_m e^{j(2\pi f_mkT + \phi_m)} + w(kT),
\end{equation}
where $w$ is a complex Gaussian noise with variance $\sigma_{w}^2$. We stress that the above model has huge applicability in \ac{rf} sensing, as it can represent the channel response in terms of reflectors distance, Doppler speed, or \ac{aoa} of the reflector, depending on the domain in which the signal is sampled (fast-time, slow-time, or spatial)~\cite{patole2017automotive, pegoraro2023rapid, liu2023sensing}. In each case, the frequency $f_m$ assumes a different meaning and it is mapped onto a propagation delay, a Doppler speed, or an angle.

\subsection{Spiking Neural Networks and spike encoding methods}\label{sec:background-snn}

\subsubsection{Spiking neuron model}

We employ the well-known \ac{lif} neuron~\cite{gerstner2014neuronal}, 
which is characterized by an internal state called \textit{membrane potential}. 
Each neuron acts as a leaky integrator of the input signal (called \textit{current}) and fires an output spike whenever its membrane potential exceeds a threshold. In the absence of input, the membrane potential decays exponentially to zero with rate $\beta$. After generating a spike, the value of the threshold is subtracted from the membrane potential.
\acp{snn} are constituted by a network of \ac{lif} neurons, typically organized into layers. The output spikes of layer $\ell$ are multiplied by learnable parameters called \textit{synaptic weights} before being fed to layer $\ell +1$. This enables optimizing the \ac{snn} using a \textit{surrogate gradient} algorithm, which approximates the standard backpropagation used in standard neural networks~\cite{eshraghian2023training}. Specifically, the non-differentiable threshold function that determines the emission of the spikes is approximated by a differentiable surrogate function during the backward pass, while in the forward pass the original thresholding is applied.

\subsubsection{Spike encoding}

The vast majority of available data for practical applications is not collected in the form of spike trains, as in the case of \ac{rf} sensing. To efficiently process this kind of data with \ac{snn}s, the need to develop \textit{spike encoding} techniques arises~\cite{auge2021survey}.
The most common spike encoding approach for sequential data is temporal contrast encoding~\cite{petro2019selection}. This class of methods keeps track of temporal changes in the signal and produces positive or negative spikes according to a thresholding criterion. 
In this work, we consider the following spike encoding algorithms:
\begin{itemize}
    \item \ac{tbr} generates a positive or negative spike when the absolute difference between two consecutive signal values (first-order difference) is higher than the threshold, and the sign of the generated spike is the sign of the difference. Given an input sequence, $x[k]$, the value of the threshold is set to
    $\mu(x') + \delta \sigma (x')$,
    where $x'$ is the sequence of first-order differences of $x$, $\mu$ and $\sigma$ denote the mean and standard deviation, respectively, and $\delta$ is a parameter of the algorithm.
    \item \ac{sf} compares the difference between consecutive values with a dynamic baseline. When the absolute difference deviates from the baseline by more than the threshold, a positive or negative spike is generated, and the baseline is updated to the current signal value. 
    \item \ac{mw} is equivalent to \ac{sf}, but the value of the baseline is set equal to the moving average of the signal at each step. Besides the threshold, this algorithm requires choosing the width of the sliding window used to compute the moving average.
\end{itemize}

\section{Learned spike encoding} \label{sec:lb-encoding}

\begin{figure}
    \centering
\captionsetup{justification=centering}
    \includegraphics[width=\columnwidth]{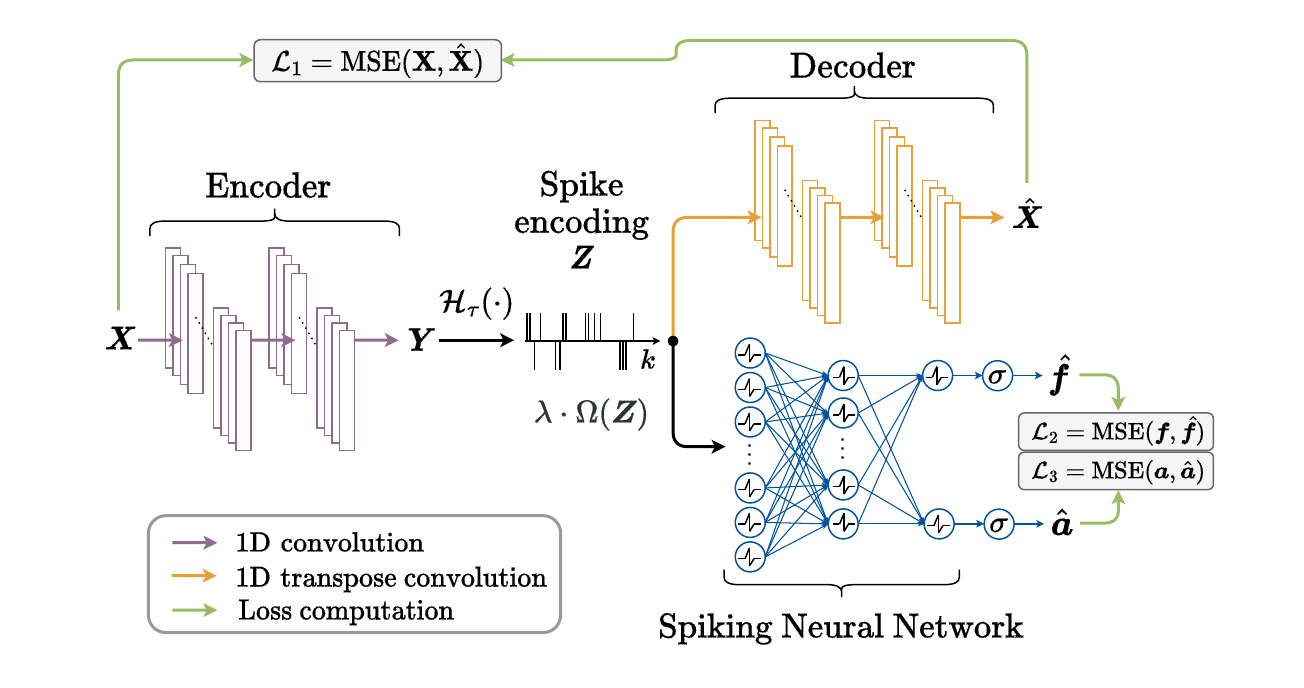}
    \caption{Block diagram of the proposed architecture.}
\label{fig:e2e_architecture}
\vspace{-1\baselineskip}
\end{figure}

Next, we describe the proposed learning-based encoding method for \ac{rf} sensing signals, shown in \fig{fig:e2e_architecture}. The neural network is composed of three main blocks: an encoder, a decoder, and an \ac{snn}, as introduced in the following.
\begin{enumerate}
\item After a preprocessing step, the \textit{encoder} block extracts salient features of the input channel response (see~\secref{sec:encoder}), which are then fed to a threshold function to produce the spike encoding (\secref{sec:spike-encoding}).
\item The \textit{decoder} block reconstructs the input channel response by approximately inverting the encoding (\secref{sec:decoder}). The network comprising the encoder and decoder blocks is referred to as \ac{cae} in the following.
\item The SNN block (\secref{sec:snn}) takes the spike encoding produced by the encoder as input and performs regression to estimate: (i)~the frequency components of the input, $f_m$, and (ii)~their amplitudes, $a_m$. By adding this block, we push the encoder to output spike encodings that preserve the spectral content of the channel.
\end{enumerate}
The whole network is jointly trained as described in \secref{sec:training}.
A detailed description of each block follows.

\subsection{Channel response preprocessing}

The input channel response is modeled as in \eq{eq:signal_model} and processed in windows of $K$ complex-valued samples, i.e., $[x[0], \dots, x[K-1]]^T$.
Before feeding it to the encoder, each processing window is transformed into a real-valued tensor, $\boldsymbol{X}$, with shape $(2, K)$, where the first dimension is the number of convolution \textit{channels} of the \ac{cae}. We call $X_c[k]$ the \mbox{$c$-th} convolution channel of $\boldsymbol{X}$ at time $k$, with $c\in \{1,2\}$.  

\subsection{Encoder}\label{sec:encoder}

The encoder takes as input the channel response processing window $\boldsymbol{X}\in \mathbb{R}^{2\times K}$ and outputs a feature representation $\boldsymbol{Y}\in \mathbb{R}^{2\times K}$. Let $E_{\xi}(\cdot): \mathbb{R}^{2\times K}\rightarrow \mathbb{R}^{2\times K}$ be the parameterized function representing the encoder, where $\xi$ is the set of weights of the encoder. Then, $\boldsymbol{Y} = E_{\xi}( \boldsymbol{X})$.

In the proposed architecture, the encoder employs $3$ one-dimensional~(1D) convolutional layers having $128$, $128$, and $2$ feature maps, respectively, and filters of length $7$ each. A stride of $1$ and zero-padding are used to preserve the input dimension. Finally, batch normalization and the hyperbolic tangent function are applied at each hidden layer, except for the last layer which has no activation function.

\subsection{Spike encoding}\label{sec:spike-encoding}

From the feature representation $\boldsymbol{Y}$, we obtain the spike encoding $\boldsymbol{Z}$, which lies in the space $\mathcal{Z}=\{-1,0,1\}^{2\times K}$. This is done by applying an element-wise \textit{hard}-threshold function, $\mathcal{H}_\tau$, to $\boldsymbol{Y}$. Using $Y_c[k], Z_c[k]$ to respectively denote the $c$-th convolution channel of $\boldsymbol{Y}$ and $\boldsymbol{Z}$ at time $k$, we have 
\begin{equation}\label{eq: customHeaviside}
    Z_c[k] = \mathcal{H}_\tau \left(Y_c[k]\right)=
    \begin{cases}
    \mathrm{sign}(Y_c[k])
    \,\, &\mathrm{if} \, |Y_c[k]|\geq \tau,\\
    \,\,0  &\mathrm{otherwise,}
    \end{cases}
\end{equation}
where $\tau >0$ is a threshold parameter. 
The two channels of $\boldsymbol{Z}$ represent the encoding of the real and imaginary part of the channel response, respectively.
It is worth noting that, although the encoder does not reduce the dimensionality of the input, it compresses it by discretizing the signal values in $\mathcal{Z}$. This leads to a loss of information that prevents the \ac{cae} from simply learning the identity function.

\textit{Remark on the use of $\mathcal{H}_\tau$:} The use of the threshold function $\mathcal{H}_\tau$ prevents a direct application of the backpropagation algorithm to train the network. Indeed, the derivative of $\mathcal{H}_\tau (Y_c[k])$ evaluates to zero everywhere, except in $Y_c[k]=\pm \tau$, where it is not defined. This causes the gradients to vanish during backpropagation.
To address this problem, we employ an \ac{ste}~\cite{bengio2013estimating}, which consists in replacing $\mathcal{H}_\tau$ with the hard hyperbolic tangent, which clamps the input value to the range $[-1,1]$, during the backward propagation of the gradients.

\subsection{Decoder} \label{sec:decoder}

We represent the decoder with a parameterized function $D_\psi(\cdot): \mathcal{Z} \rightarrow  \mathbb{R}^{2\times K}$, where $\psi$ is the set of weights of the decoder. The reconstructed channel response is $\hat{\boldsymbol{X}} = D_{\psi}( \boldsymbol{Z})$.
The decoder is composed of $3$ \emph{transposed} 1D convolutional layers with $128, 128$, and $2$ filters, respectively. Filter size, stride, padding, batch normalization, and activation function are the same as those used in the encoder, except for the last layer which implements the sigmoid function.

\subsection{Spiking Neural Network}\label{sec:snn}

The \ac{snn} processes the input spike encoding, $\boldsymbol{Z}$, to regress the frequencies and amplitudes of the sinusoidal components in the input signal, denoted by $\hat{\bm{f}}=[\hat{f}_1,\dots,\hat{f}_M]^T\in\mathbb{R}^M$ and $\hat{\boldsymbol{a}}=[\hat{a}_1,\dots,\hat{a}_M]^T\in\mathbb{R}^M$, respectively. 
Denoting by $\gamma$ the set of parameters of the \ac{snn}, we represent the network with the function \mbox{$S_\gamma: \mathcal{Z} \rightarrow  \mathbb{R}^{M} \times \mathbb{R}^{M}$}, such that $(\hat{\bm{f}}, \hat{\boldsymbol{a}}) = S_{\gamma}( \boldsymbol{Z})$.
To do so, the \ac{snn} interprets $\boldsymbol{Z}$ as a $K$ timesteps-long sequence of $2$-dimensional vectors, each containing the real and imaginary spike components. Specifically, we define the input vector to the \ac{snn} at time $k$ as $\boldsymbol{z}[k] \triangleq [Z_1[k], Z_2[k]]^T$. 

The \ac{snn} is composed of \ac{lif} layers.
Each layer takes as input the spike train produced by the previous layer and behaves as a leaky integrator until a firing threshold is reached. At layer $\ell$, denote by 
$\boldsymbol{u}_{\ell}[k]$ the vector of membrane potentials at time $k$, by $\boldsymbol{\theta}_{\ell}$ the vector containing the firing thresholds of each neuron, by $\boldsymbol{W}_{\ell}$ the learnable weights matrix, and by $\boldsymbol{\beta}_{\ell}$ the vector of membrane potential decay constants~\cite{eshraghian2023training}. 
The input-output equation of the $\ell$-th \ac{lif} layer is 
\begin{equation}\label{eq:lif-layer}
    \boldsymbol{u}_{\ell}[k] = \boldsymbol{\beta}_{\ell} \boldsymbol{u}_{\ell}[k-1] + \boldsymbol{W}_{\ell}\boldsymbol{s}_{\ell-1}[k] - \boldsymbol{\theta}_{\ell}\boldsymbol{s}_{\ell}[k-1],
\end{equation}
where vector $\boldsymbol{s}_{\ell}[k]$ is the output spike vector of layer $\ell$ at time $k$. 
Note that for the first layer we set $\boldsymbol{s}_{0}[k] = \boldsymbol{z}[k]$. 

Our \ac{snn} architecture comprises an input layer with $128$ neurons, a hidden layer with $64$ neurons, and \textit{two} parallel output layers, which we denote by $\mathcal{O}_f$ and $\mathcal{O}_a$, with $5$ neurons each. The output layers are specifically designed to reconstruct the channel frequencies and the amplitudes, respectively. The output spikes from the hidden layer are processed in parallel by $\mathcal{O}_f$ and $\mathcal{O}_a$ and then the membrane potential of each output layer is passed through the sigmoid function to obtain $\hat{\bm{f}}$ and $\hat{\bm{a}}$, respectively.

The membrane potential decay rates, $\boldsymbol{\beta}_{\ell}$, and the firing thresholds, $\boldsymbol{\theta}_{\ell}$, are learned during training, hence the set of learnable parameters of the \ac{snn}, $\gamma$, contains $\boldsymbol{W}_{\ell}, \boldsymbol{\beta}_{\ell}, \boldsymbol{\theta}_{\ell} $ for the four layers.

\subsection{Network training and hyperparameter optimization}\label{sec:training}
In this section, we describe the training process of the learned spike encoding network.

\subsubsection{Loss function and sparsity regularization}\label{sec:regularization}
The network is trained end-to-end through the minimization of a loss function $\mathcal{L}$ which depends on the outputs of the decoder and the \ac{snn}, as well as on the spike encoding itself.
Specifically, let $\boldsymbol{f}$ and $\boldsymbol{a}$ be the vectors of the true frequency and amplitude components of the channel response, respectively, with components $f_m, a_m$. 
We define $3$ reconstruction loss terms using the \ac{mse} as 
\begin{equation}
\mathcal{L}_1\triangleq \mathrm{MSE}(\boldsymbol{X}, \hat{\boldsymbol{X}})= \frac{1}{2K}\sum_{c=1}^2\sum_{k=0}^{K-1}(X_c[k]-\hat{X}_c[k])^2,
\end{equation}
$\mathcal{L}_2 \triangleq\mathrm{MSE}(\boldsymbol{f}, \hat{\bm{f}})= \sum_{m=1}^{M} (f_m-\hat{f}_m)^2/M$, and $\mathcal{L}_3\triangleq\mathrm{MSE}(\boldsymbol{a}, \hat{\boldsymbol{a}})= \sum_{m=1}^{M} (a_m-\hat{a}_m)^2 /M$.
The frequencies, amplitudes, and values of $\boldsymbol{X}$ are normalized in the range $[0,1]$, as usual practice with neural networks. 
Note that this implies that all the error terms also lie in $[0, 1]$.

Additionally, a regularization term $\Omega (\boldsymbol{Z})$, which depends on the number of spikes, is added to the loss to penalize non-sparse spike trains. 
The use of this regularization is motivated by the objective of obtaining \emph{sparse} spike trains since the number of processed spikes is tied to the energy consumption of the system.
Recalling that $\boldsymbol{Z}\in \{-1,0,1\}^{2\times K}$, we define
\begin{equation}
    \Omega (\boldsymbol{Z}) \triangleq \frac{1}{2K}\sum_{c=1}^2 \sum_{k=0}^{K-1} |Z_{c}[k]|,
\end{equation}
where the absolute value is due to the presence of negative spikes.  We use a regularization coefficient $\lambda\in [0,1]$, that allows us to tune the importance of the regularization term in the loss and gain control of the energy content of the spike encoding. Larger values of $\lambda$ lead to higher sparsity (lower energy consumption), while $\lambda=0$ means no regularization is applied (higher energy consumption). 
The total loss function $\mathcal{L}$ is obtained as $\mathcal{L} = \mathcal{L}_1 + \mathcal{L}_2 + \mathcal{L}_3 + \lambda\,\Omega (\boldsymbol{Z})$.

\subsubsection{Training}
\noindent We train the network in batches of $256$ samples, using backpropagation with the Adam optimizer with learning rate $\eta=10^{-3}$. 
In computing the gradients of the \ac{snn} parameters, we use the \textit{surrogate gradient} approach to avoid the non-differentiability of the \ac{lif} layer~\cite{eshraghian2023training}. Specifically, we adopt the arctangent approximation of the firing threshold function during the backward pass.
The training process is stopped upon convergence of the loss on a validation set.

\subsubsection{Hyperparameter optimization}

We optimize the network hyperparameters by performing a greedy search, using the loss on a validation set as the optimization objective. Notably, we obtain that the best threshold function for the spike encoding layer is $\tau =0.1$.
 
\section{Simulation and numerical results}\label{sec:results}

In this section, we compare the results obtained by the proposed architecture, referred to as \ac{lse} in the following, and the encoding methods from the literature, presented in~\secref{sec:background-snn}. All the code is developed in Python, using the PyTorch and the snnTorch frameworks~\cite{eshraghian2023training}.

\subsection{Synthetic dataset generation}

We create a synthetic dataset using the model presented in \secref{sec:signal-model}. To simulate realistic \ac{rf} channel responses, we used typical parameters of \ac{rf} sensing systems for Doppler spectrum estimation in the $60$~GHz unlicensed band, see e.g. \cite{pegoraro2023rapid}. In this setting, frequencies $f_m$ represent Doppler shifts induced by the motion of objects in the environment. This is just one of the possible applications of our proposed encoding algorithm, which is general enough to be also applied to range or \ac{aoa} estimation. 

Following \cite{pegoraro2023rapid}, we set $T=0.27$~ms, and $K=64$, while the carrier frequency is $f_c = 60$~GHz. This choice of $T$ leads to a maximum observable Doppler velocity of $v_{\rm max}=c/(4f_cT)\approx4.48$~m/s, where $c$ is the speed of light.

To generate channel responses with a variable number of sinusoidal components, we set $M=5$, to be intended as the \textit{maximum} number of components in the synthetic data. Then, channel responses with fewer than $M$ components are obtained by setting to $0$ the corresponding amplitudes $a_m$. We generate $3,000$ windows for each possible number of sinusoidal components in $\{1, \dots, M\}$, for a total of $15,000$. The dataset is randomly split into training, validation, and test sets with the proportion $75$:$15$:$10$. 
We generate each channel response window by applying the following pipeline: 
 \begin{enumerate}
     \item the frequency $f_m$ of each sinusoidal component is randomly generated in the interval $[0, 1/(2T)]$;
     \item the phase $\phi_m$ of each component is randomly generated in the interval $[0, 2\pi)$;
     \item the amplitudes are randomly generated and normalized with respect to the maximum one so that it equals $1$;
     \item the complex Gaussian noise is generated such that the \ac{snr} belongs to the set $\{5, 10, 15, 20\}$~dB.
 \end{enumerate}

\subsection{Comparison with state-of-the-art algorithms}

In this section, we compare our learned encoding algorithm to the state-of-the-art methods introduced in \secref{sec:background-snn}.
Recall that the quality of the final encoding provided by the \ac{tbr}, \ac{sf}, and \ac{mw} algorithms heavily depends on the choice of the input parameters. Hence, for each algorithm, we performed a grid search to find the values of the parameters that minimize $\mathrm{MSE}(\boldsymbol{X}, \hat{\boldsymbol{X}})$.
For each algorithm, the specific search space and final optimal values are reported in \tab{tab:gridsearch}.

\begin{table}[t!]
    \caption{Optimization of \ac{tbr}, \ac{sf}, and \ac{mw} parameters.}
    \label{tab:gridsearch}
    \begin{center}
    \begin{tabular}{l c c c}
    \toprule
        \textbf{Method} & \textbf{Search interval}  & \textbf{Step value} & \textbf{Optimal value} \\ 
        \midrule 
        \textbf{TBR} (factor $\delta$) & $[0.001, 0.01]$ & 0.001 & 0.005 \\ 
        \textbf{SF} (threshold) & $[0.05, 0.3]$ & 0.05 & 0.2 \\ 
        \textbf{MW} (threshold) & $[0.01, 0.6]$ & 0.05 & 0.06 \\ 
        \textbf{MW} (window) & $[2, 4]$ & 1 & 3 \\
        \bottomrule
    \end{tabular}
    \end{center}
\end{table}

\begin{table}[t!]
    \caption{Performance comparison (normalized reconstruction errors and average sparsity) between our method (LSE) and threshold-based methods. We report the mean and standard deviation on the test set.}
    \label{tab:results}
    \begin{center}
    \begin{tabular}{l c c c}
    \toprule
        \textbf{Method} & \textbf{Rec. RMSE}  & \textbf{DFT mag. RMSE} & \textbf{Sparsity} \\ 
        \midrule 
        \textbf{TBR} \cite{petro2019selection} & 0.374 $\pm$ 0.075  & 0.043 $\pm$ 0.010 & 0.015 \\ 
        \textbf{SF} \cite{petro2019selection} & 0.222 $\pm$ 0.044& 0.029 $\pm$ 0.009 & 0.433 \\ 
        \textbf{MW} \cite{petro2019selection} & 0.262 $\pm$ 0.059 & 0.039 $\pm$ 0.011 & 0.202 \\ 
        \textbf{LSE} & 0.133 $\pm$ 0.028 & 0.017 $\pm$ 0.004 & 0.736 \\
        \bottomrule
    \end{tabular}
    \end{center}
    \vspace{-1\baselineskip}
\end{table}

\begin{figure}[t!]
\centerline{\includegraphics[width=\columnwidth]{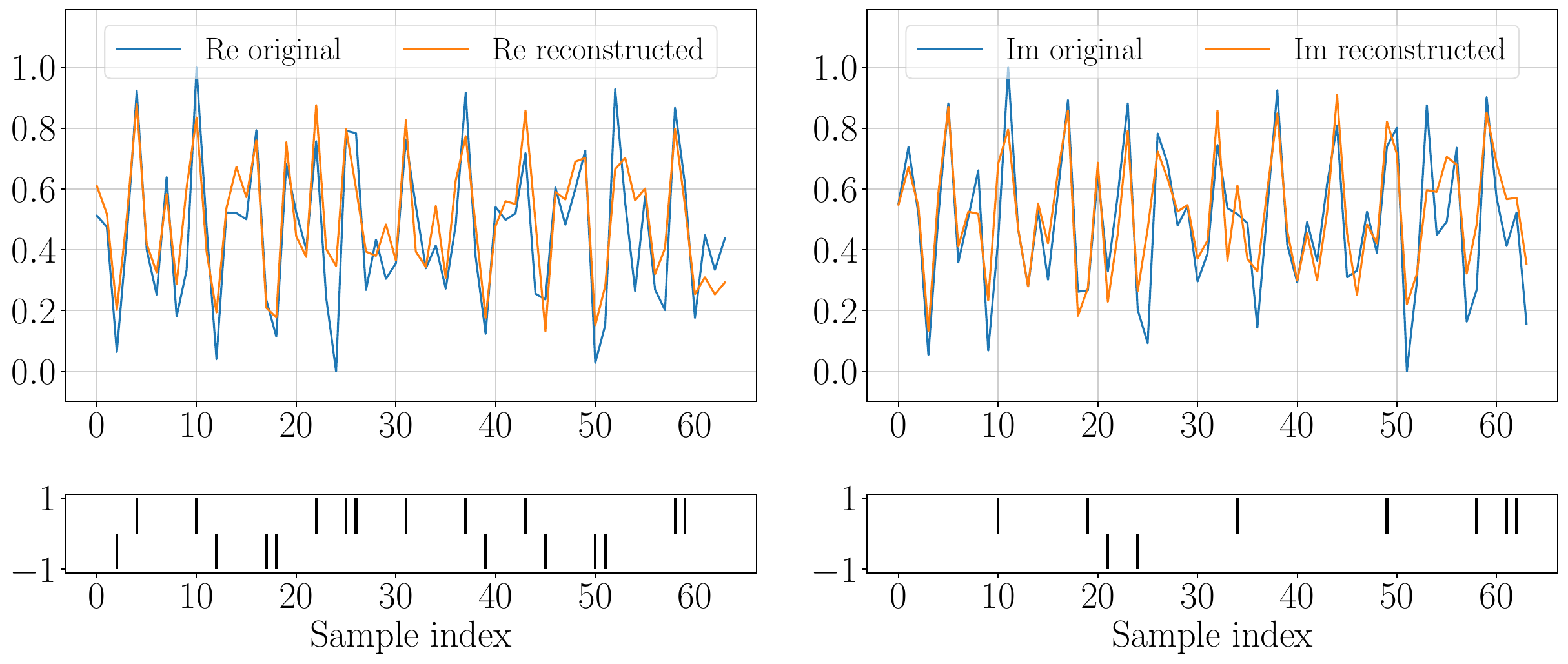}}
\caption{Qualitative comparison between the reconstructed channel and the original one (top), and the learned spike encoding (bottom), where the real part of the channel (Re) is shown on the left and the imaginary one (Im) on the right.}
\label{fig:reconstruction}
\vspace{-1\baselineskip}
\end{figure}

\subsubsection{Reconstruction error}
To evaluate the quality of the reconstructed channel, we adopt the \textit{per-window} \ac{rmse} computed over the test set, obtained as \mbox{$\mathrm{RMSE}(\boldsymbol{X}, \hat{\boldsymbol{X}}) = \sqrt{\mathrm{MSE}(\boldsymbol{X}, \hat{\boldsymbol{X}})}$}. All errors are then averaged over all the windows of the test set, to obtain the mean reconstruction error.
As shown in \tab{tab:results}, \ac{lse} achieves a remarkably low reconstruction error compared to 
other methods. 
Indeed, temporal contrast techniques are hindered by the use of a fixed threshold to encode the signal, which prevents the model from capturing fine-grained, dynamic channel features. 
An example of a test window reconstructed by LSE, along with the obtained spike train, is shown in~\fig{fig:reconstruction}. 

\subsubsection{Spectral components preservation}

We evaluate and compare the encoding algorithms on their capability of preserving the channel spectral content by computing the per-window \ac{rmse} of the reconstructed channel response \ac{dft} magnitude with respect to the original one. As shown in \tab{tab:results}, our learned encoding demonstrates superior performance due to its higher representation power. As an additional result, we report that our method also achieves good results in the estimation of the channel frequency components through the \ac{snn}, achieving an average per-window \ac{rmse} of $0.1$ for the frequencies and $0.16$ for the amplitudes. 

\subsubsection{Sparsity of the encoding}

\begin{figure}[t!]
\centerline{\includegraphics[width=\columnwidth]{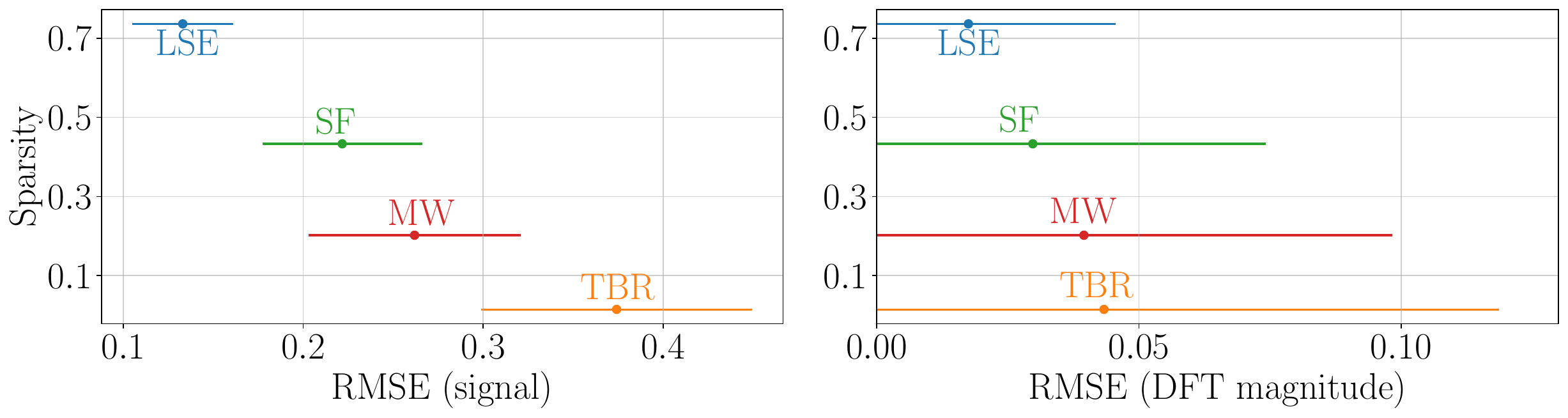}}
\caption{Reconstruction RMSE for different spike encoding sparsity levels of the channel response (left) and its DFT magnitude (right).}
\label{fig:sparsity}
\vspace{-1\baselineskip}
\end{figure}

\begin{figure}[t!]
\centerline{\includegraphics[width=5.5cm]{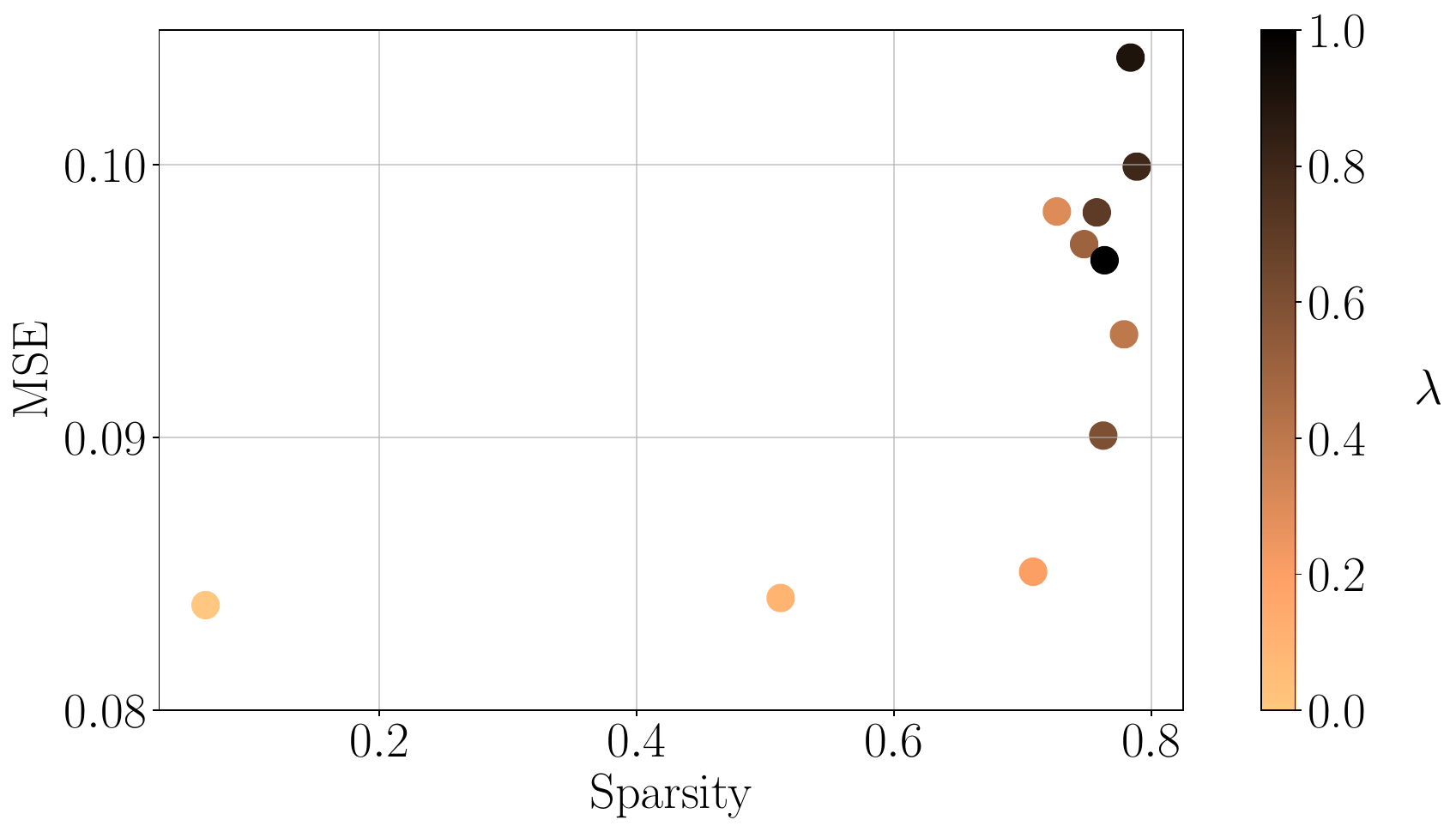}}
\caption{Sparsity level for different values of $\lambda$ using the learned spike encoding.}
\label{fig:sparsity_lse}
\vspace{-1\baselineskip}
\end{figure}

We define the sparsity of a spike encoding as the fraction of zeros in $\boldsymbol{Z}$, i.e. $\rho(\boldsymbol{Z}) = 1 - \Omega(\boldsymbol{Z})$.
$\rho(\boldsymbol{Z})$ directly correlates with the efficiency of the \ac{snn} used for processing the spike trains. 
The sparsity of each method is then obtained by averaging the sparsity of all the encoded test windows, as reported in \tab{tab:results}. 
\ac{lse} exhibits $70\%$ higher sparsity than the best algorithm from the literature, which is obtained by \ac{sf} ($0.73$ vs. $0.43$). This demonstrates its efficiency in representing key features with minimal redundancy (this result is obtained using $\lambda = 0.2$). Other standard encoding methods, especially \ac{tbr}, produce denser spike trains. \fig{fig:sparsity} highlights the trade-off between encoding sparsity and reconstruction accuracy. Our \ac{lse} offers a substantial improvement over existing approaches in encoding \ac{rf} sensing data. Moreover, the proposed approach allows directly \textit{controlling} the sparsity-accuracy trade-off by tuning the regularization parameter $\lambda$.
This is shown in \fig{fig:sparsity_lse}, where we plot the sparsity and the corresponding validation reconstruction \ac{mse} for different values of $\lambda$ (averaged over $14$ training with different initialization). Higher $\lambda$ leads to higher sparsity (lower energy consumption) but implies performance degradation (higher \ac{mse}).

\subsection{Robustness to noise}

\begin{figure}[t!]
\centerline{\includegraphics[width=\columnwidth]{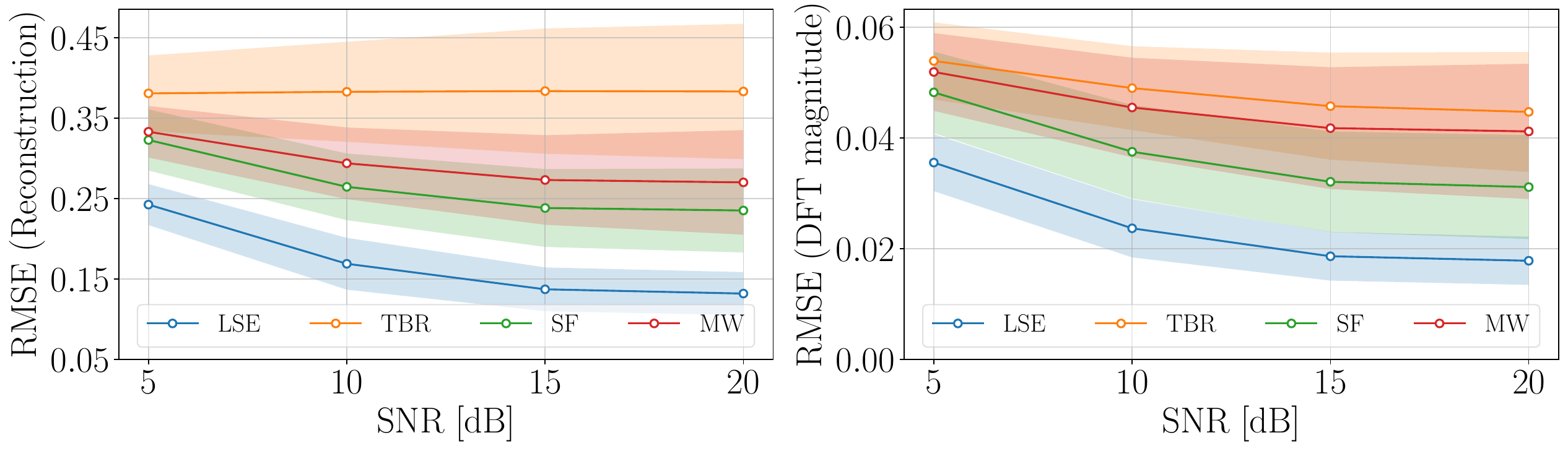}}
\caption{Robustness to noise when reconstructing the channel response (left) and its \ac{dft} magnitude (right). Shaded areas represent the standard deviation.}
\label{fig:robustness}
\vspace{-1\baselineskip}
\end{figure}

Lastly, we analyze the robustness of each method to noise. To this end, we generated multiple test sets, with $10,000$ samples each, with different \ac{snr} values ($5, 10, 15$, and $20~\mathrm{dB}$), following the pipeline outlined in \secref{sec:signal-model}. Hence, for each test set and encoding method, we computed the reconstruction and \ac{dft} magnitude errors. The results are shown in \fig{fig:robustness}. \ac{lse} obtains the lowest errors and improves the most at high \ac{snr}. Notably, \ac{lse} also has the lowest standard deviation across different \ac{snr} values, which makes it more reliable.

\section{Concluding remarks}

In this work, we have designed and implemented the first neural network that learns a \emph{tailored} spike encoding for \ac{rf} channel responses in \ac{rf} sensing applications. 
To this end, we proposed a \ac{cae} to obtain the spike encoding of the input data, enforcing sparsity by penalizing dense spike trains. 
In addition, we used an \ac{snn} block to perform a regression on both the frequency and the amplitude components of the input data, to push the spike train to preserve the original spectral content. The proposed network is trained end-to-end on a synthetic dataset. Our results show that it learns to produce a better spike encoding than existing thresholding-based methods, both in terms of fidelity to the spectral content of the original signal and sparsity of the spike trains.
While our primary objective is to shift from traditional \ac{dl} architectures to \ac{snn} for energy efficiency reasons, it is important to highlight the compactness of the proposed \ac{dl}-based encoder architecture (fewer than $120$~K parameters and less than $2$~MB of size in memory), which ensures a lower inference time, compared to that of the \ac{snn}, and allows its deployment on low-memory edge devices.
Moreover, the proposed technique allows controlling the \textit{performance-energy} trade-off through a regularization parameter that trades information content of the spike encoding with its sparsity, which is directly connected to the network energy consumption. This provides high flexibility in adapting it to different applications and hardware implementations. 

Future developments of this work include: (i)~an evaluation of the trained encoder on real \ac{rf} sensing data, (ii)~an in-depth analysis of the energy consumption of the encoding during inference, (iii)~the possibility of integrating the spike encoding step directly into the \ac{snn}, and (iv)~the training of the whole network to autonomously recognize the different sinusoidal components without having to set a maximum number a priori.

\bibliography{references}
\bibliographystyle{ieeetr}

\end{document}

%% file: acronyms.tex
\acrodef{aoa}[AoA]{Angle of Arrival}
\acrodef{cae}[CAE]{Convolutional Autoencoder}
\acrodef{dl}[DL]{Deep Learning}
\acrodef{dft}[DFT]{Discrete Fourier Transform}
\acrodef{fmcw}[FMCW]{Frequency-Modulated Continuous-Wave}
\acrodef{isac}[ISAC]{Integrated Sensing and Communication}
\acrodef{lif}[LIF]{Leaky Integrate and Fire}
\acrodef{lse}[LSE]{Learned Spike Encoding}
\acrodef{mw}[MW]{Moving-Window}
\acrodef{ofdm}[OFDM]{Orthogonal Frequency Division Multiplexing}
\acrodef{snn}[SNN]{Spiking Neural Network}
\acrodef{sf}[SF]{Step-Forward}
\acrodef{ste}[STE]{Straight-Through Estimator} 
\acrodef{snr}[SNR]{signal-to-noise ratio} 
\acrodef{rf}[RF]{Radio Frequency}
\acrodef{tbr}[TBR]{Threshold-Based Representation}
\acrodef{uwb}[UWB]{Ultra-Wideband}
\acrodef{rmse}[RMSE]{Root Mean Squared Error}
\acrodef{conv1d}[1D-Conv.]{one-dimensional convolutional}
\acrodef{mse}[MSE]{Mean Squared Error}